\input epsf.tex

\documentstyle[12pt]{article}

\newcommand{\be}{\begin{equation}}   \newcommand{\ee}{\end{equation}}
\newcommand{\bear}{\begin{eqnarray}}
\newcommand{\eear}{\end{eqnarray}}
\newcommand{\ba}{\begin{array}}      \newcommand{\ea}{\end{array}}

\begin{document}
\pagestyle{empty}
\begin{titlepage}

\vspace*{-8mm}
\noindent 
\makebox[11.5cm][l]{BUHEP-96-32} February 4, 1997\\

\vspace{2.cm}
\begin{center}
  {\LARGE {\bf  Vacuum  Tunneling by Cosmic Strings }}\\
\vspace{42pt}
%{\large 
Indranil Dasgupta \footnote{e-mail address:
dgupta@budoe.bu.edu}

\vspace*{0.5cm}

 \ \ Department of Physics, Boston University \\
{590 Commonwealth Avenue, Boston, MA 02215, USA}

\vskip 3.4cm
\end{center}
\baselineskip=18pt

\begin{abstract}

{\normalsize
We consider vacuum tunneling of a new kind where the false vacua are
not translationally invariant, but have topological
defects that break some of their translational symmetries. In the
particular case where the topological defects are cosmic strings,
we show the existence of an $O(2) \times O(2)$ symmetric bounce
configuration that allows a semi-classical estimate of the
rate of cosmic string induced tunneling. A method of reduction is
then suggested for simplifying the computation of the bounce action.
Some phenomenological applications are described.
}

\end{abstract}

\vfill
\end{titlepage}

\baselineskip=18pt  
\pagestyle{plain}
\setcounter{page}{1}

%%%%%%%%%%%%%%%%%%%%%%%%%%%%%%%%%%%%%%%%%%%%%%%%%%%%%%%%%%%%%%%%
\section{Introduction}

%%%%%%%%%%%%%%%%%%%%%%%%%%%%%%%%%%%%%%%%%%%%%%%%%%%%%%%%%%%%%%%%%%%%%
\label{sec:intro}

First order phase transitions in field theories may sometimes take place
through the nucleation of bubbles seeded by massive excitations \cite
{seed}. Of the several distinct processes that can lead to the
seeding, a particularly interesting one is due to the
instability of topological defects (like cosmic strings and monopoles)
with respect to spontaneous dissociation. 
The phenomenon is best understood by an example. Consider
the $SU(5)$ symmetric GUT theory. The sequence of phase
transitions $SU(5) \to SU(4) \times U(1) \to SU(3) \times SU(2) \times
U(1)$ occurs as the universe cools from temperatures of $10^{15}$
GeV. At the end of the first phase transition $SU(4) \times U(1)$
monopoles are formed. These monopoles may have a structure such that at
their cores the 
vacuum expectation values (VEV) of the scalar fields responsible for the second
phase transition lie near the minimum of the effective potential of the
$SU(3) \times SU(2) \times U(1)$ phase. As the temperature drops to the
point where the $SU(4) \times U(1)$ phase is metastable compared to the 
$SU(3) \times SU(2) \times U(1)$ phase, the monopoles become
unstable if it is energetically favourable for their cores
to expand. In other words
the monopoles themselves become bubbles of true vacuum inside a sea
of false vacuum. It was shown in ref. \cite {s1} that this phenomenon
may precede the nucleation of the usual bubbles in the false vacuum and the
phase transition may take place essentially through bubbles seeded by
monopoles.

An analogous situation is the case where the topological
defects remain classically stable as
the supercooling proceeds but may decay by quantum tunneling. 
The result is the same
seeding effect discussed above. However seeded bubbles compete with
unseeded bubbles on the same footing; both are quantum (or thermal) 
processes and
suppressed due to the usual barrier penetration (or Boltzman) factors and 
one must compute the rate of
nucleation of the two kinds of bubbles 
to decide whether
the seeding has an appreciable effect or not.

The phenomenon is interesting in theories where the false vacuum
is so long lived that the phase transition never takes place and the
universe remains trapped in the supercooled phase forever. In
the minimal supersymmetric model (MSSM), for instance, it may be possible
that the true vacuum is charge and color breaking (CCB) \cite {ccb} 
and the existing, viable vacuum is only a false vacuum. 
Recently it has been shown that such viable false vacua can exist
in other low energy supersymmetry breaking models \cite {ddr} that also
permit the existence of 
$U(1)$ cosmic strings. In all these cases, the decay rate of the viable false
vacuum provides interesting constraints on 
the model. However, the
existence of topological defects in the false vacua of these theories is
also 
quite generic and independent constraints can be obtained by
considering the nucleation rate of seeded and unseeded bubbles. While 
the tunneling associated with unseeded bubbles is well understood
semiclassically and the rate can be found by computing the action of
$O(4)$ invariant bounce configurations of the underlying Euclidean field
theory, a similar semiclassical treatment is needed in the case of the
seeded bubbles for a useful estimate of the corresponding
tunneling rates. 

In this paper we treat a local region of
space containing a topological defect as a translationally non-invariant
(non-homogeneous) 
false vacuum which, classically, corresponds to a well defined
local minimum of the
energy functional of the theory. We show that there are new bounce
configurations (or instantons) associated with these non-homogeneous vacua 
that permit a semiclassical
understanding of the tunneling seeded by the defect.
In particular there is a bounce with a large symmetry that appears as
the global minimum of the action in an open space of field
configurations. 
We then devise simple numerical 
methods to compute a lower bound on the corresponding semiclassical tunneling
rates. Although the new bounce configurations are 
more complicated than the usual bounces, we use a 
method of ``reduction'' to simplify the computation to 
the point where it is
not much harder than the $O(4)$ symmetric 
computations for the usual bounces \cite {dg}. 

We restrict 
ourselves to the zero temperature tunneling, since that is the most
interesting case for theories where the false vacuum is viable and 
required to be very long lived. However, generalizing to the finite
temperature case is straightforward and the computation of the free
energy of the critical bubble is analogous to the computation of the
bounce action in a 3 dimensional Euclidean field theory. 
Our treatment is also  
applicable to all types of topological defects except global texture
(see \cite {pv} and references therein for a discussion of different
types of defects), but for brevity we limit
ourselves to the case of cosmic strings only. In this case 
the effect of the
tunneling is the random appearence of bead-like bubbles along the length
of the cosmic string, which captures the essential feature of tunneling
seeded by extended defects. Note that 
cosmic strings also happen
to be the most interesting type of defects from the point of view of
cosmology (\cite {av} and references therein). 

The paper is organized as follows. In section 2 we prove
the existence of the new type of bounce solutions and show that they
appear as global minima of the action in an open space of field
configurations. 
In section 3 we indicate some phenomenological applications and effects of the 
seeded tunneling phenomenon. 
Finally in section 4 we suggest a method
of ``reduction'' that makes the estimation of the bounce action feasible
even in complicated field theory models. 

\section {The Bounce as a Global Minimum of Action}

The semiclassical vacuum tunneling rate 
was computed
in ref. \cite {cc} by a saddle point evaluation of the Euclidean
path integral. In the case of a
theory with a single real scalar field $\phi$ with a false vacuum at
$\phi \equiv \phi_f$, the transition
probability per
unit volume per unit time was found to be:
\be
{\Gamma \over V}= \left ({S_E [\overline{\phi}] \over 2 \pi \hbar}\right ) ^2
{\rm {exp}}\left ( {-S_E[ \overline{\phi} ]\over \hbar}\right ) 
\left| {{\rm {det}}
^{\prime} [-\partial ^2+U^{\prime \prime}(\overline {\phi})] \over
{\rm {det}}[-\partial ^2+U^{\prime \prime}(\phi_f)]} \right|^{-1/2}
\times
\left (1 +O(\hbar)\right ) \, , 
\label {transition}
\ee
where $S_E[ \overline{\phi} ]$ is the Euclidean action for the so called 
`bounce' configuration $\overline {\phi}$ and $\partial ^2 = \partial
_{\mu} \partial _{\mu}$. The prime on the
determinant indicates that the zero eigenvalues are omitted.

We will use the Abelian Higgs model with a single complex scalar $\Phi$
to make our arguments concrete. To develop the formalism we start with 
the Euclidean action:
\be \label {gaugeaction}
S_E= \int d^4x \left [ {1 \over 4g^2} F_{\mu \nu}F^{\mu \nu}+{1\over 2}|D
\Phi|^2 + U(\Phi) \right ] \, ,
\ee
where $D_{\mu}\Phi= (\partial _{\mu}-iA_{\mu})\Phi$ and $F_{\mu \nu}=
\partial _{\mu} A_{\nu}-\partial _{\nu} A_{\mu}$.

Let us suppose that the potential
$U(\Phi)$ has a global minimum at $\Phi_t$ and
a local minimum at $\Phi_f \ne 0$. The false vacuum at $\Phi_f$ may
permit the existence of cosmic strings. 
A false vacuum with a single cosmic string of unit winding
number is the lowest energy state with the following boundary
condition for the VEV
 of the scalar field (in some appropriate gauge),
\be \label {stringone}
{\rm {lim}}\, _{x,y \to \infty} \Phi (r, \theta) = |\Phi_f|\, {\rm {exp}}
\, (i\theta)
\ee
where $r^2=x^2+y^2, {\rm {tan}}
 (\theta)=y/x$. The cosmic string runs parallel
to the $z$ axis and is a static solution of the equations of motion. 
Apart from the cylindrical symmetry, the solution is invariant under
translations in the $z$ direction. 

The translationally invariant and homogeneous 
false vacuum at $\Phi \equiv \Phi_f$ is
metastable and its decay rate is determined by (\ref {transition}). 
The corresponding 
bounce configuration $\overline {\Phi} ^0$ is likely to be an $O(4)$
symmetric field configuration and the tunneling rate corresponds to the
nucleation rate of unseeded bubbles. 
However the lowest energy state with an infinite cosmic string is also 
a local minimum of the energy functional and one can 
estimate the tunneling rate
associated with its decay. We will treat a region of space with an 
isolated straight cosmic string as a classical false vacuum which is
simply non-homogeneous. 
The formalism behind 
equation (\ref {transition}) applies. The only important 
difference is that  the field
configuration  at the false vacuum is a space dependent function which is
not invariant under translations in $x$ and $y$ directions. The bounce
configuration in this case is likely to be more complicated and will
certainly lack the $O(4)$ symmetry of the usual bounces. The
corresponding tunneling rate will be interpreted as the nucleation rate of
critical bubbles along the length of the cosmic string.

When the false vacuum is chosen to have a single infinite 
cosmic string with the above properties, the Euclidean action for the
new bounces must be computed with the energy density of the false vacuum
normalized to zero. Let us denote the field configuration corresponding
to the local
minimum of energy with a single cosmic string by $\Phi^0, A^0$
and the corresponding covariant derivative and field strength 
by $D^0\Phi^0$ and $F^0_{\mu \nu}$.
The appropriately normalized Euclidean action is then
(dropping the subscript on $S$):
\be \label {normalaction}
S= K_1 + K_2 + P + F_1 + F_2 + F_3 \, , 
\ee
with $K_1= \int d^4x {1 \over 2} \left [|{D_z \Phi}|^2 +|{D_t \Phi}|^2
\right]; \,
K_2= \int d^4x {1 \over 2} \left [|{D_x \Phi}|^2 +|{D_y \Phi}|^2
- |{D_x \Phi^0}|^2 -|{D_y \Phi^0}|^2 \right]; \, \\
P =\int d^4x \left [U(\Phi)-U(\Phi^0) \right ]; \,
F_1= \int d^4x {1 \over 4g^2} \left [F_{xy}^2 - (F^0_{xy})^2 \right ]; \, 
F_2= \int d^4x {1 \over 4g^2} \left [F_{zx}^2+F_{tx}^2+F_{zy}^2+F_{ty}^2
\right ] $ and $F_3 =\int d^4x {1 \over 4g^2}\left [F_{zt}^2 \right ]$. 
Here we have used the result that, 
$\Phi^0$ and $ A^0$ have cylindrical symmetry around the $z$ axis
with $A_z^0 = A_t^0 =
K_1^0=F^0_{zx}=F^0_{zy}=F^0_{tx}=F^0_{ty}=0$. 
We will refer to this solution of the equations of motion 
as the trivial solution $G^0$. 
The regularized quantities
$K_2, F_1$ and $P$ are finite in the neighbourhood of $G^0$.

The search for stationary points of the action is greatly simplified by
imposing symmetries on the field configurations. We will impose 
rotational symmetries in the 
$x,y$ as well as in the $z,t$ plane. Introducing the polar coordinates 
$R^2 = z^2 + t^2$, ${\rm {tan }}(\gamma ) = z/t$, $r^2 = x^2 + y^2$ and 
${\rm {tan}}(\theta) = y/x$ we write
$\Phi(x,y,z,t) = a(r,R)
{\rm {exp}} (i \theta)$, $A_{\theta} = v(r,R)/r$ and $A_r=A_R=A_{\gamma}=0$
where $a$ and $v$
are real. The action reduces to:
\be
S = 2\pi^2 \int \int RdR\; rdr
 \left [ {1\over 2} \left ( |\partial _R a|^2 + |\partial _r a|^2 
 \right )+ {1\over
  2}(1-v)^2 {a^2 \over r^2} + {1 \over 4g^2 r^2} (\partial _r v)^2 + U(a)
- {\cal {L}}_0\right ] \; ,
\label {numersone}
\ee
where the counterterm ${\cal {L}}_0$ is obtained by replacing $a$ and $v$
by the two functions 
$a_0(r)$ and $v_0(r)$ corresponding to the trivial solution $G^0$. 
The above ansatz for the fields allows us 
to set $A_r, A_R, A_{\gamma}=0$. Variations in $\Phi$ and $A$ that do
not preserve the ansatz contribute only positive definite terms to
the action and therefore the  ansatz minimizes the action with respect
to those variations (this can also be seen as a consequence of the
principle of symmetric criticality \cite {palais}).
Consequently a stationary point of the action in (\ref
{numersone}) is a stationary point of the full action (\ref {normalaction}). 
The ansatz also implies that $F_2 = F_3 = 0$. 
In the rest of the paper we always take $A_t, A_z, A_r, F_2, F_3
\equiv 0$ and enforce the $O(2) \times O(2)$ symmetry described above. 
Also, by fixing the phase of $\Phi$ in the above
fashion we fix the gauge completely.

The trivial solution $G^0$ has the same $O(2) \times O(2)$ symmetry and 
is obtained by finding the minimum energy
solution with the boundary conditions (\ref {stringone}). 
The energy functional is given by
\be \label {energytwo}
E = 2 \pi\int dz \int rdr \left [ {1\over 2}|\partial _r a|^2 + {1\over
  2}(1-v)^2 {a^2 \over r^2} + {1 \over 4g^2 r^2} (\partial _r v)^2 + U(a)
\right ].
\ee
Minimizing $ E$ gives the equations of motion:
\bear 
{\partial ^2 a \over \partial r^2} = {\partial U(a) \over \partial a} +
(1-v)^2 {a \over r^2} - {1 \over r} {\partial a \over \partial r} \\
\nonumber
 {\partial ^2 v \over \partial r^2} =  {1 \over r} {\partial v \over
  \partial r} - {2 g^2a^2} (1-v).
\label {eqmotion}
\eear
$G^0$ is the solution $a_0(r), v_0(r)$ of the above equations
that satisfies the boundary conditions:
$a_0(0)= 0, a_0(\infty) = |\Phi_f|, v_0(0)= 0, v_0(\infty) = 1$. 

Note that with the normalization of (\ref {normalaction}) 
the action of $G^0$ is zero and it is a local minimum of $S$.
In contrast, 
a bounce solution in this case must have the following properties:\\
(i) It starts at Euclidean
time $t=-\infty$ from the static 
cosmic string solution and ends at Euclidean
time $t=+\infty$ at the same solution but is not given everywhere by the
the  trivial time translation of the static string solution. \\
(ii) The solution, which extremizes the action, is a saddle point of
the action with a single direction of instability, i.e with a single
negative eigenvalue of the Hessian ${\partial ^2 S \over \partial
 [\Phi, A]^2}$.\\
(iii) All velocities
($d\Phi /dt$ and $dA_{\mu}/dt$) are zero for a 
temporal slice at
some finite time (which we can choose to be at $t=0$). The field
configuration at $t=0$ is often called the turning point of the bounce.

The direction of instability associated with the bounce configuration is 
intimately related to scale transformations of the fields. 
For a systematic exploration of these
transformations it will be useful to define a one parameter
deformation of the action (\ref {normalaction}) 
in the following manner (after dropping $F_2$
and $F_3$): 
\be
S[\eta] = K_1[\eta]+ K_2 + P + F_1 \; ;
\label {etadeforms}
\ee
where $K_2, P, F_1$ 
are as defined before, $K_1[\eta]$ is
a deformation of $K_1$ defined by :
\be
K_1[\eta] = \int d^4x {1 \over 2} \left [|{D_z \Phi}|^{2-\eta} +
|{D_t \Phi}|^{2-\eta} \right]\; .
\label {etadeformk}
\ee
In the limit $\eta \to 0$ we recover the usual Abelian
Higgs model (with the constraint $A_z, A_r, F_2, F_3 \equiv 0$).

Under the scale transformation $(z,t) \to (\lambda z, \lambda t)$, with
$\lambda$ being a positive number, the terms in the deformed action
$S[\eta]$ transform as:
$K_1[\eta]  \to  \lambda ^{\eta} K_1[\eta] \; ; 
(K_2, P, F_1)  \to  \lambda ^2 (K_2, P, F_1) $. 
Under these transformations 
the configuration $G^0$ is left invariant while the finiteness
of the action is preserved in the 
neighbourhood of $G^0$.
An extremum of the deformed 
action must be invariant under the scale transformation.
Therefore at a stationary point of the deformed 
action we have,
\be 
\label {lambdatwo}
{ \delta S[\eta] \over \delta \lambda} \Big |_{\lambda =1} =
\eta \, (K_1[\eta])+ 2\left [ {P}+{K_2}+{F_1} \right ]  =0.
\label {lambdathree}
\ee
The second derivative of $S[\eta]$ at this point is:
\be 
{ \delta ^2 S[\eta] \over \delta \lambda ^2} \Big |_{\lambda = 1} 
= \eta (\eta - 1) K_1[\eta] + 
2\left [ {P} + {K_2} +{F_1} \right ] \; .
\label {lambdafour} 
\ee
Since $K_1[\eta] > 0 $ for any non-trivial field configuration, 
the above quantity is 
negative for a non-trivial extremum of $S[\eta]$ if $\, 2 >\eta > 0$. 
We will define the quantity $V= \left [{P}+ {K_2} +{F_1} \right ] $ and
prove the following statement. 

{\bf Statement 1:} {\it {A bounce configuration 
exists in the Abelian Higgs model with the deformed action provided $2 > \eta
> 0$. The bounce is the global minimum of the action in an open space of
field configurations.}}

{\bf Proof:} 
Let the space of all smooth 
field configurations having the boundary condition given
by property (i) of the bounce, satisfying (\ref {lambdathree})
and having the $O(2) \times O(2)$ symmetry 
$\Phi (x,y,z,t) = |\Phi| (r,R) {\rm {exp}} (i \theta)$ 
be $C$. The trivial solution $G^0$ lies in C. Every point in $C$
automatically satisfies properties (i) and (iii) of the bounce. We need
to show that there is a point in $C$ that also satisfies property (ii).

We can split $C$
into two parts $C=C_0 \oplus C_1$ where $C_0$ is a connected space in
$C$ containing $G^0$. In fact $C_0$ consists of a single point $G^0$ 
since there
exists a ball around $G^0$ such that all field configurations
contained in the ball except $G^0$
have $V>0$.
This follows from observing that at the
false vacuum the energy density is minimized and therefore $G^0$ 
is a local minimum of $S[\eta]$ with $V=0$. However all nontrivial
points on $C$ have $K_1[\eta]> 0$ and therefore by (\ref {lambdathree})
$V< 0$.

The space $C_1$ is not empty since one can always construct a
non-trivial field configuration that satisfies (\ref {lambdathree}) and 
the correct boundary conditions. By construction,
the global minimum $G$ in $C_1$ is
is an extremum of the action with respect to all
allowed variations.
By (\ref {lambdafour}) it also has a single direction of instability
corresponding to the scale transformations $(z,t) \to \lambda (z,t)$. 
Therefore it
has property (ii) of the bounce. {\bf Q.E.D.}

The limit $\eta = 0$ can be approached in the following way. 
By definition, $G$ is the global minimum of $C_1$. First we show that
$G$ remains a maximum with respect to variations in a direction ``orthogonal'' to
$C_1$ even when $\eta = 0$. The easiest way to see this is to consider a
curve ${\cal C}[p]$ with $p \in [-1,1]$ in the space of all field
configurations with the right boundary conditions and
passing through $G$ (${\cal C}[0]=G$) in a direction ``orthogonal'' to the
space $C_1$. For $2> \eta > 0$ one can draw this curve so that one end
of the curve is at $G^0$ (${\cal C}[-1]=G^0$) and the other end
approaches the point $G^1$ where $\Phi \equiv \Phi_t$ (${\cal C}[1]\to
G^1$). Note that the space $C_1$ forms a boundary between the
neighbourhoods of $G^0$ and $G^1$. This is because $V[G^0]=0$ and
$V[G^1] \to -\infty$, therefore the curve  ${\cal C}[p]$ must 
intersect the codimension one surface $C_1$ where $V<0$. 
The curve can be drawn so that  $S\left [{\cal C}[p]\right ]\le S[G] $
for all $p \ne 0$. As $\eta \to 0$, this property of the curve remains
true and $G$ remains a maximum of the action with respect to the variation that
corresponds to moving along this curve. In this sense $G$ remains a
saddle point even when $\eta = 0$. In principle the negative 
eigenvalue of the Hessian
${\partial ^2 S \over \partial [\Phi, A]^2}$ may approach zero as $\eta
\to 0$. Nevertheless, the zero eigenvalue can not be generic since there
is no underlying symmetry that requires a zero eigenvalue. Phrased
differently, all eigenvalues of the Hessian are likely to be of the
order of $m^2$ where $m$ is some characteristic mass scale in the
theory. An eigenvalue can not be made zero without fine tuning the
parameters of the theory, unless there is a symmetry to protect its
smallness. Hence, for generic theories the limit $\eta \to 0$ can be
safely taken. 

Although the above discussion has been carried out in the context of an
Abelian Higgs model with a single Higgs, everything said so far applies
to the case of multiple Higgs fields. The generalization to the
non-Abelian case is straightforward too; one simply needs to reduce the
action in the non-Abelian case to an action resembling the Abelian model
by dropping all components of the gauge field except one corresponding
to a broken generator that generates a non-contractible loop on the
vacuum manifold.

So far we have proven the existence of a bounce with a large symmetry
($O(2) \times O(2)$)
that that also happens to be the global minimum of the action in a
simple space of field configurations. As we shall see, the latter property 
makes it quite simple to bound the action of this bounce from above.
If this bounce is the bounce with
the least action then its action determines the tunneling rate of
equation (\ref {transition}). If there is a bounce with a smaller action (and a
lesser symmetry) then $G$ provides only an upper bound on the least
bounce action and consequently a lower bound on the tunneling rate. 
This bound itself may be useful in phenomenological applications and in
section 4 we suggest simple numerical techniques for obtaining
this bound.

\section {Phenomenological Applications}

The formal expression for
the tunneling rate is similar to equation (\ref {transition}).
However we must drop the zero modes from the primed
determinant.
There are 
just two zero modes of the bounce
solution coming from translations in the $z$ and $t$ directions. 
To leading order in $\hbar$ we have:
\be
\gamma = {\Gamma \over L}= \left ({S [\overline{\Phi}] \over 2 \pi \hbar }\right ) 
{\rm {exp}}\left ( {-S[ \overline{\Phi} ]\over \hbar}\right ) 
\left| {{\rm {det}}
^{\prime} [-\partial ^2+U^{\prime \prime}(\overline {\Phi},\overline {A}
)] \over
{\rm {det}}[-\partial ^2+U^{\prime \prime}(\Phi^0, A^0)]} \right|^{-1/2} \, ,
\label {newtransition}
\ee
where $S$ is the action of the string induced 
bounce $[\overline {\Phi}, \overline {A}]$ (if and when it exists). 
Note that the configurations $\Phi^0, A^0$ are space dependent.
Corresponding to the two zero modes there are only two factors of $\left
({S [\overline{\Phi}] \over 2 \pi \hbar }\right )^{1/2}$ on the R.H.S
\cite {cc}. The ratio of the determinants in (\ref {newtransition}) 
now has the dimensions of
mass squared and the transition rate is to be interpreted
as the probability of transition per unit time {\it {per unit length of the
string}}.
Collecting all non-computable parts into a single prefactor $\cal {A}$ 
we can rewrite the above equation as:
\be
\gamma = {\cal {A}}\,
{\rm {exp}} (-S[\overline {\Phi},\overline {A}]/\hbar) \; .
\label {stringtrans}
\ee
To a good approximation 
$\cal {A}$ can be replaced by the characteristic squared 
mass scale of the theory and only the exponent needs to be determined
carefully. 

Phenomenologically the tunneling rate
obtained from (\ref {stringtrans}) is useful in two distinct
situations. 
The first is the case when the
theory has a viable false vacuum and one requires that 
the phase transition should never
occur. The second case is a GUT model at high temperatures where one
would like to calculate the high temperature vacuum decay rate and see
if the string induced bubbles can percolate and complete the phase
transition. We will briefly discuss both cases.

Consider the case of the viable false vacuum first.
In this case, the false vacuum should have a lifetime longer than the
age of the universe. A constraint is therefore obtained from the zero
temperature tunneling rate.
Multiplying the transition rate by the total
world sheet area of the cosmic strings, one gets the
expected number of bubbles seeded by 
cosmic strings in a given space time. 
Most of the contribution to the string world sheet area 
comes from later times when the string network has entered a scale
invariant distribution \cite {av}. During these times about $80\% $
of the string length is contained in one long string per
horizon. The total world sheet area available for bubble nucleation is
therefore $ \sim H^2$ where $H$ is the present horizon distance. 
The false vacuum would be regarded unstable if ${\Gamma
  \over L} \times H^2 \ge 1$.
If the mass scale of the theory is greater than the 
electroweak scale
then this condition translates into the following bound on the bounce
action:
\be \label {sboundone}
S[\overline {\Phi},\overline {A}] \le 200 \, \hbar.
\ee
The same arguments repeated for unseeded bubbles 
yields the condition:
\be \label {sboundtwo}
S[\overline {\Phi ^0},\overline {A ^0}] \le 400 \, \hbar.
\ee
Here $[\overline {\Phi ^0}, \overline {A^0}]$ is the ususal 
$O(4)$ invariant bounce and
we have used the condition that ${\Gamma \over V} \times H^4
\ge 1$. Notice that the tunneling rate is expressed as the number of bubbles
per unit time per unit {\it {volume}}
 and correspondingly one uses the total
space time {\it {volume}} available for nucleating the bubbles.

Cosmologically, there is no inflation in the false (but viable)
vaccum and the bubbles, once formed, eventually percolate. The stability
of the false vacuum therefore demands that both conditions (\ref
{sboundone}) and (\ref {sboundtwo}) be untrue. 
The significance of this lies in the fact that a-priori the bounce
actions $S[\overline {\Phi},\overline {A}]$ and 
$S[\overline {\Phi}^0,\overline {A}^0]$
are not related in any simple manner
and constraints obtained on the parameter space of
the model by (\ref {sboundone}) are independent and may be stronger than
the constraints implied by (\ref {sboundtwo}). 
The simplest way to see this is to note that it is possible to have
theories where the cosmic strings are classically unstable 
\cite {s2}. 
By a continuous deformation of the parameters of 
a theory one can go from a regime of classically unstable strings to
metastable strings. Clearly there is a continuous family of 
theories in between where
the bounce action in (\ref {sboundone}) is arbitrarily small compared to
the bounce action in (\ref {sboundtwo}) and the string induced
tunneling is the dominant tunneling effect.

Now consider the case when the viable vacuum is the true vacuum. 
The interesting case arises when the universe gets trapped in the false
vacuum and supercools. Then
the phase transition must be completed by seeded and unseeded bubbles in
an inflating universe which (in comoving coordinates) 
is described by the Robertson-Walker metric:
\be 
d\tau ^2 = dt^2 - R^2(t)\left [ {dr^2 \over 1-kr^2} + r^2 d\Omega ^2
\right ] \; , 
\label {rw}
\ee
where $k=-1, 0$ or $+1$ depending on whether the universe is open, flat
or closed respectively. The scale factor $R(t)$ grows according to the
equation:
\be 
\left ({{\dot {R}} \over R}\right )^2 = {8 \pi \over 3 M_p^2} \rho - {k \over R^2} \; ,
\label {scale}
\ee
where $M_p= 1.2 \times 10^{19}$ GeV is the Planck mass. The energy
density $\rho$ is given by $\rho = \rho_0 + \left ( {\pi ^2 \over 30 }
\right ) \eta T^4$ where $\rho_0 $ is the energy density of the false
vacuum, $\eta$ is the number of effectively massless degrees of freedom
and $T$ is the temperature. When the temperature is below the critical
temperature $T_c$ where the false vacuum becomes metastable, the energy
density of the vacuum $\rho_0$ makes the dominant contribution to $\rho$.
Approximating the energy density
$\rho$ by the vacuum energy density $\rho_0$ we obtain in the period
of inflation: ${{\dot {R}} \over R} \approx \chi \rho_0$,
and the scale factor grows exponentially with time ($R \propto {\rm {exp}}
(\chi t) $) where $\chi = \left ({8 \pi \rho_0 \over 3 M_p^2} \right
)^{1/2}$. With adiabatic expansion we also have: $RT={\rm {constant}}$, 
so the
temperature falls exponentially with time.

In an inflating universe, the nucleation rate of unseeded bubbles grows
as the volume of the universe ($\sim R^3$) while the same rate for
seeded bubbles grows only as the length of the cosmic strings ($ \sim
R$). Nevertheless, the seeded bubbles can play a significant and even
the dominant role in the phase transition. To see this 
consider the probability that an
arbitrary point remains in the supercooled false vacuum at time
$t$. This was
found in ref. \cite {wg1} to be:
\be
p(t) = {\rm {exp}} \left [ - \int ^t _{t_c} dt^{\prime} n(t^{\prime})
V(t,t^{\prime}) \right ] \; ,
\label {pt}
\ee
where $V(t,t^{\prime}) = {4\pi \over 3} \left [ \int ^t _ {t^{\prime}}
{d t^{\prime \prime} \over R(t^{\prime \prime})} \right ] ^3$ is the
coordinate volume at time $t$ of a bubble formed at time $t^{\prime}$
and $n(t^{\prime})dt^{\prime}$ 
is the number of bubbles formed per unit coordinate volume 
between the times $t^{\prime}$ and
$t^{\prime} + dt^{\prime}$ and $t_c$ is the time corresponding to the
temperature $T_c$.
If the tunneling rates associated with seeded and unseeded
bubbles are respectively $\gamma_0$ and $\gamma$ repectively, we
can write:
\be
n(t) = \gamma_0(t)R^3(t) + \gamma(t)\sigma R(t)R^2(t_c) \; ,
\label {nt}
\ee
where $\sigma $ is the total string length per unit coordinate volume at
time $t_c$. 

Both $\gamma_0$ and $\gamma$ are the finite temperature tunneling rates
which depend on time (or temperature) and can be evaluated by computing
the bounce action in the 3 dimensional Euclidean space
using the finite temperature effective potential. The tunneling rates
will however
be greater or equal to their zero temperature values $\gamma(0)$
and $\gamma_0(0)$. 
With the approximations made above we can
re-write equation (\ref {pt}) as a function of temperature.
\be
p(T) \le {\rm {exp}} \left [-d \int ^{T_c}_{T} {dT_1 \over T_1}
 \left ({\gamma_0(0) \over T_1^3 }+ {\sigma
\gamma(0) \over T_1 T_c^2} \right )
 \left ( \int ^{T_1} _{T}dT_2 \right )^3 \right ] \; ,
\label {pT}
\ee
with $d={4 \pi \over 3 \chi^4}$. One can look at the above expression in
two limiting cases. If the seeded bubbles have little effect on the
phase transition we can drop the terms multiplying $\gamma$ from (\ref
{pT}). Then in the limit $T \to 0$ we have:
\be
p(T \to 0) \le {\rm {exp}} \left [-d\gamma_0(0) {\rm {log}}\left ({T_c \over
T}\right ) \right ] \; .
\label {gammaz}
\ee
If $d \gamma_0(0)$ is of the order of unity or greater, a fast phase
transition is accomplished. 
For small values of $d \gamma_0(0)$  ($d \gamma_0(0) \le 10^{-3}$ may be
sufficiently small) there
may be no percolation or the resulting universe may be too
inhomogeneous if reheating is achieved through bubble collisions \cite
{wg2}. Therefore viability of a model requires that either the
transition be fast with only a mild inflation or the first order phase
transition is followed by a slow roll-over of the inflaton as in new
inflation. 

The effect of the seeded bubbles can be seen by putting
$\gamma (0) = 0$ in (\ref {pT}). Then we get:
\be
p(T \to 0) \le {\rm {exp}} \left [-d\sigma \gamma(0)  \right ] \; .
\label {gamma}
\ee
In this case the fraction of unconverted space approaches a non-zero
value ($p \ne 0$) as $T \to 0$. However if $d \sigma \gamma (0) \ge 1$ then the
phase transition is fast and the
resulting universe is homogeneous with the temperature at the end of the
phase transition being of the order of $T_c$. Comparison of (\ref
{gammaz}) and (\ref {gamma}) shows that for $\sigma \gamma (0) 
\ge \gamma_0(0)$ the seeded bubbles make a significant
contribution to the phase transition. The value of $\sigma$ 
depends on the initial distribution of the cosmic strings. There are two
extreme cases. If the temperature at which the strings are formed is
close to $T_C$ then the string network is at the friction dominated
period when the inflation begins and $\sigma \sim T_c^2$. If on the
other hand the strings are formed at a much higher temperature than
$T_c$ then they would have a scaling distribution at $T_c$ and we would
expect about one long string per horizon. This implies that $\sigma \sim
T_c^4 /M_{pl}^2$. The characteristic mass scale associated with the
tunneling rates $\gamma_0$ and $\gamma$ is likely to be $T_c$. Therefore
we can write: $\gamma_0(0) \sim T_c^4 {\rm {exp}}(- {S}^0/\hbar)$
and $\gamma (0) \sim T_c^2 {\rm {exp}}(- {S}/\hbar)$, where
${S}$ and $\overline {S}^0$ are the corresponding zero
temperature bounce actions. Using
these expressions the condition for the seeded bubbles to be significant
in the transition may be stated as: $S -\alpha  \le
{S}^0$, where $\alpha$ lies in the interval $[0, {\rm
{log}}(T_c^2 / M_{pl}^2)]$. Given the fact that the two bounce actions
are independent of each other, in any realistic model
the above condition is as likely to be
satisfied as not.

\section {Computing The Bounce Action}

In this section we 
present a simple technique 
for actually computing the 
the zero temperature tunneling rate.
The basic idea is similar to the suggestions of 
ref. \cite {dg} and ref. \cite {kusenko}.
Instead of looking for a saddle point of 
the action in the space
of all configurations, one looks for the global minimum of the action in
the reduced space $C_1$. 
The restriction to the reduced space can be done by Lagrange multipliers. We may
define
an improved action $\overline {S}$ by:
\be 
\overline {S} = S + \Sigma _i M_i \, .
\label {impaction}
\ee
The extra terms $M_i$ added to the action are of the form $c_jB_j^n$
where $c_j$ is a Lagrange multiplier, $n$ is a positive number and $B_j$
are functionals of the fields that vanish on the reduced space.
For instance, $\overline {S} = S + c_1 V^2 $ has local minima
at the trivial solution $G^0$ as well as $G$. By choosing $c_1$ 
large one can essentially force an iterative minimization to take
place along a path on the surface $C_1$. Clearly the minimum one may 
obtain is
either $G$ or some other local minimum $G^{\prime}$ 
of $S$ in $C_1$. In either case
one reaches a bounce. 
However, since $C_1$ is an infinite dimensional
space, in practice one can not arrive at the point
$G$ or even a local minium $G^{\prime}$ 
through finite methods. Indeed, the result of the search is likely
to be a point which lies close to $C_1$ but may not even be on $C_1$. 
Nevertheless, by performing a small scale
transformation $(z,t) \to \lambda (z,t)$, 
one can make the final point of the search lie exactly on $C_1$.
Thus irrespective of the accuracy of the numerical
method, the search always succeeds in finding an upper bound on the
action of the bounce $G$ which sets a useful lower bound on the
tunneling rate. 

Another advantage of this method is that it admits extreme
simplifications. 
Since one can at best hope to find an upper bound on
the action of $G$, the computation can be simplified enormously at an
early stage by making the search space small. The only
required element of the computation is that condition $V=0$
be enforced so that the
result of the search is a point on $C_1$. 
We will use the example of the Abelian Higgs 
model to illustrate a simplified search of this kind where the problem
will be reduced to solving an ordinary differential equation in a single
variable. (As mentioned before, the non-Abelian case is reducible to the
Abelian one). 
We start with the action given by (\ref {numersone}). The most interesting case 
arises when the 
potential $U(\Phi)$ has a local minimum at $|\Phi| = \Phi_f \ne 0$ and a
global minimum at $\Phi = 0$. This is, for instance, the case in the
MSSM if the true vacuum is viable and the false vacuum breaks
hypercharge. The cosmic strings in the false vacuum may be metastable or
even classically unstable with respect to an expansion of their cores. 
In the true vacuum, they simply do not exist. 

The first step is to determine the functions $a_0(r)$ and $v_0(r)$ 
numerically by
minimizing the energy functional (\ref {energytwo}) or by 
solving the equations (\ref {eqmotion}) numerically. Next, consider the
field configuration $
a(r,R)  = 
a_0(r-r^{\prime}(R)) \; {\rm for }\; r \ge  r^{\prime}(R)\, ;\;  a(r,R) =
0 \; {\rm for }\; r  <  r^{\prime}(R) \, ;\\
v(r,R)  =
v_0(r-r^{\prime}(R)) \; {\rm for }\; r \ge  r^{\prime}(R)\, ; 
\; v(r,R) = 0 \; {\rm for }\; r  <  r^{\prime}(R) \, ;$
where $r^{\prime}(R)$ is a real positive function of $R$ satisfying
$r^{\prime}(R)|_{R \to \infty} = 0$ and ${dr^{\prime} \over dR}|_{R=0}
=0$. 
The field 
configuration consists of a string that nucleates a bubble of true
vacuum in its 
``core'' at some $t < 0$. The bubble expands 
as one approaches $t=0$ and then it contracts and vanishes at
some $t > 0$. For a given value of $z$ and $t$, the cross section of the
bubble in the $x,y$ plane is a disc of radius $r^{\prime}$.

\vspace*{-1.3cm}
\centerline {\hbox {
{\epsfxsize=2.5in\epsfbox{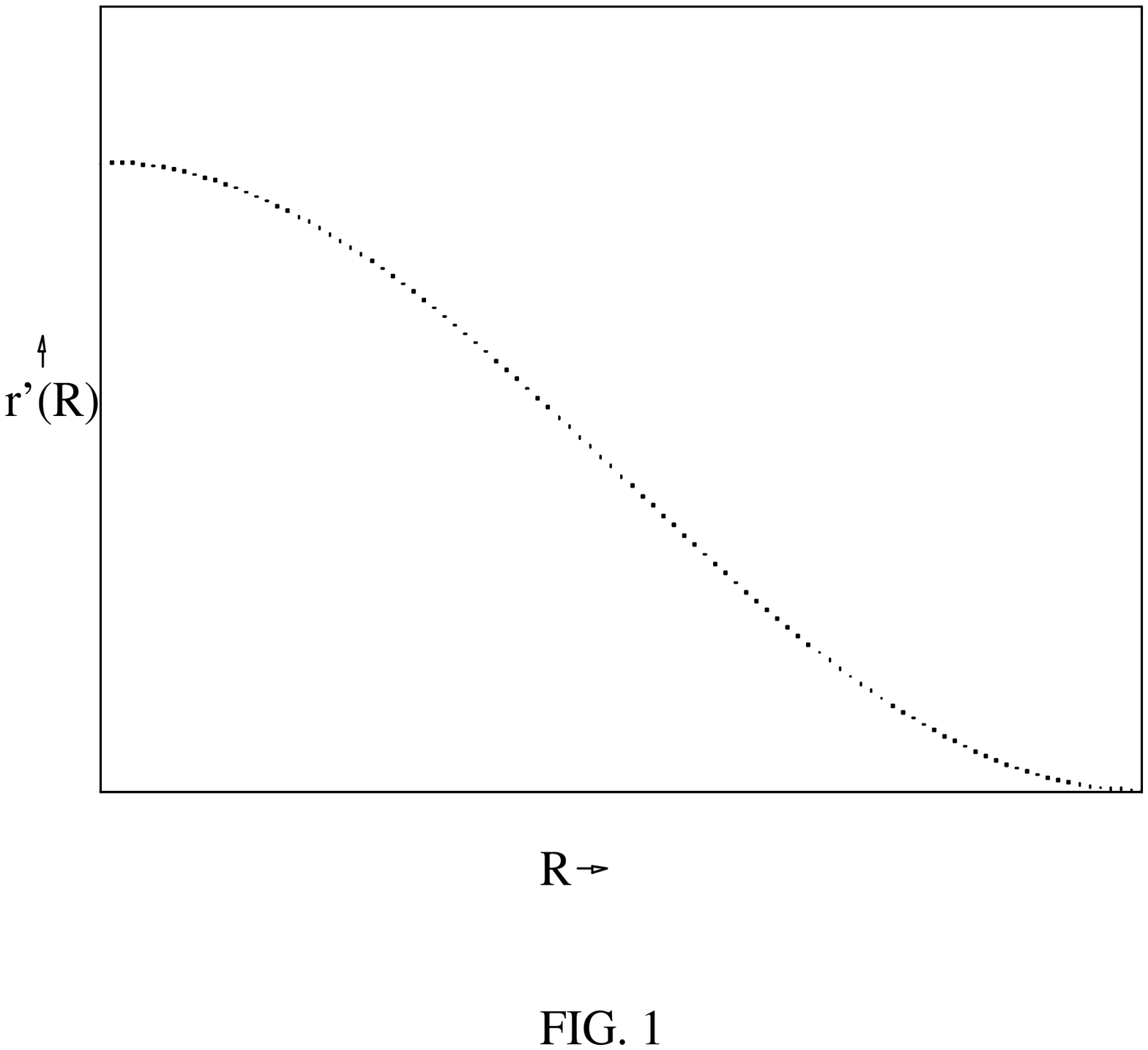}} \hspace {2cm}
{\epsfxsize=2.5in\epsfbox{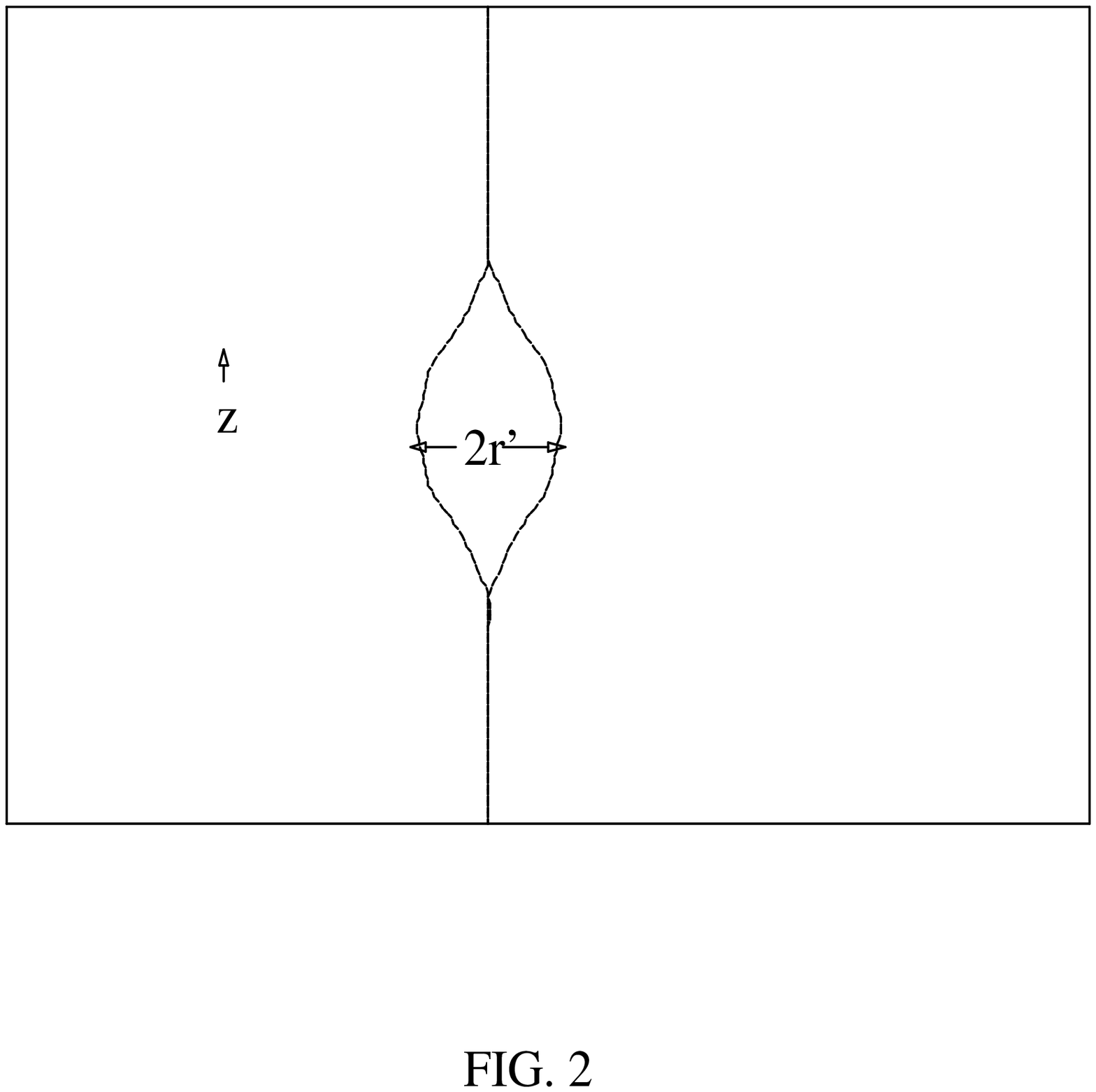}}
}}
\noindent
\makebox[0.8in][l]{\hspace{2ex} FIG. 1.}
\parbox[t]{4.8in}{ {\small  {A typical form for the function
      $r^{\prime}(R)$. The function goes to zero as $R \to \infty$. } }}

\noindent
\makebox[0.8in][l]{\hspace{2ex} FIG. 2.}
\parbox[t]{4.8in}{ {\small  {A schematic diagram for the string ``core''
after the nucleation of the bubble. The radius of the bubble in a plane
perpendicular to the $z$ axis is $r^{\prime}$. } }}
\vskip 0.5cm

The above construction gives us a space of
field configurations satisfying the correct boundary conditions and 
depending  on a single real function $r^{\prime}$ of $R$. Restricted 
to this space of field configurations, the
action is simply a functional of $r^{\prime}(R)$ which can be thought of
as the 
Lagrangian of a particle with coordinate $r^{\prime}$
integrated over time $R$.
We have:
\bear
S &=& 2 \pi^2 \int R\, dR
\left [ (L_1 + r^{\prime}L_2)  (\partial _R r^{\prime})^2 
+  (L_3 + r^{\prime}L_4) + {U(0) \over 4} r^{\prime 2} - {\cal {L}}_0
\right ] \; , 
\label {numerstwo}
\eear
where $ L= \left [{a_0^2 \over r^2} (1-v_0)^2 + {1 \over 2g^2r^2}
(\partial _r v_0)^2 \right ]; \; 
L_1= \int r\, dr (\partial _r a_0)^2;\; L_2= \int  dr (\partial
_r a_0)^2; \; L_3= \int r\, dr L; \;$ and $ L_4 = \int  dr L$.
The integrals in $r$ must be done numerically
with the known functions
$a_0$ and $v_0$. 
Then the bounce in the reduced theory is obtained by solving the
equation of motion implied by (\ref {numerstwo}) which is an ordinary
differential equation.
This is exactly
the same computation that one needs to do to evaluate the tunneling rate
in the usual translationally invariant case for a theory with a single
real scalar field. It can be easily done by the shooting method 
that has been described elsewhere \cite
{cc, dg, kusenko}. 

Note that at the extremum, 
the action is extremized with respect to scale transformations of $R$ which
correspond to 
scale transformations in the $z,t$ plane. Thus the extremum
automatically 
lies on the surface $C_1$ and gives an upper bound on the
action of $G$. In this case there is no further need to include Lagrange
multiplier terms as in (\ref {impaction}) to force the result to lie on
the surface $C_1$. This feature is typical of
reductions of this kind where the final action depends on a single real
function \cite {dg}.

\section{ Conclusions}

We have considered vacuum tunneling from non-homogeneous
false vacua. A summary of our results is as follows:

(i) There are new tunneling phenomena associated with topological
defects in a false vacuum. The rate of vacuum
tunneling through bubbles seeded by defects is independent of 
(and can be arbitrarily large compared to)
the tunneling rate through bubbles that are not seeded by the defects.

(ii) We have identified a bounce configuration (and call it $G$) which 
allows a semi-classical estimation 
of the tunneling rate.

(iii) It is possible to search for $G$ numerically, by using simple
minimization of a function on the lattice. We suggest a method of
reduction by which one can do the search over a reduced space and 
find an upper bound on the action at $G$. The corresponding tunneling
rate is a lower bound on the actual tunneling rate, and may be useful
in placing constraints on the parameter space of realistic 
field theory models.

\vskip 1.0cm
{\centerline {\bf {Acknowledgements}}}
I am grateful to Claudio Rebbi for many useful conversations.
I would like to thank Bogdan Dobrescu, Robert Brandenberger, 
Eric Weinberg, So Young Pi, Ryan Rohm and Paul Steinhardt
for valuable ideas and suggestions. 
This work was supported by the Department of Energy under the grant
DE-FG02-91ER40676.

\vfil

\end {document}